\documentclass[preprint,amsmath,amssymb,prc,aps]{revtex4-2}
\begin{document}

\title{{\Large Partial-wave analysis to determine the spin of $\Xi(1690)^{-}$ and $\Xi(1820)^{-}$ produced in $\overline{p}p$ annihilation}}

\author{Deepak Pachattu}

\email{deepakpn@goa.bits-pilani.ac.in}
\affiliation{Department of Physics, BITS Pilani K K Birla Goa Campus, Zuarinagar 403726, Goa, India}

\begin{abstract}
Recently \cite{panda} the PANDA collaboration studied the feasibility of determining the spin and parity of the $\Xi(1690)^{-}$and $\Xi(1820)^{-}$ resonances in the $\Lambda K^-$ system produced in $\bar{p}p$ collisions via the reaction channel $\overline{p} p \rightarrow \bar{\Xi}^{+} \Lambda K^{-}$. The purpose of this contribution is to study these reactions using a model-independent irreducible tensor formalism developed earlier. This study leads us to identify the partial-wave amplitude which would be zero if $\Xi(1690)^{-}$ or $\Xi(1820)^{-}$ had spin-1/2, provided these resonances are produced at threshold (s-wave production).
\end{abstract}

\maketitle
\section{Formalism}
\subsection{$\bar{p} p \rightarrow \bar{\Xi}^{+} \Xi^{*-}$} 
The spin structure \cite{gr-msv,npa} for the reaction of the type $a+A\to b+B$ involving particles/reactants $a,A$ and products $b,B$ with respective spins $s_a,s_A$ and $s_b,s_B$  can be studied by defining a matrix $\mathcal{M}$ of dimension $\left(2 s_b+1\right)\left(2 s_B+1\right) \times\left(2 s_a+1\right)\left(2 s_A+1\right)$ in the spin-space of the participants. The matrix elements are related to the on-energy-shell transition matrix $\mathcal{T}$ through
$$
\left\langle s_b \mu_b s_B \mu_B|\mathcal{M}| s_a \mu_a s_A \mu_A\right\rangle=\sqrt{\frac{2 \pi D}{v}}\left\langle s_b \mu_b s_B \mu_B ; \overrightarrow{p_2}|\mathcal{T}| \overrightarrow{p_i} ; s_a \mu_a s_A \mu_A\right\rangle
$$
where $\overrightarrow{p_2}$ and $\overrightarrow{p_i}$ are the final and initial c.m. momenta respectively. The density of final states and the magnitude of the relative velocity in the initial state are denoted by $D$ and $v$ respectively. Introducing channel spins $s_i$ and $s$ and partial wave expansions in the entrance and exit channels, we have,
\begin{equation}
\begin{aligned}
\label{partial}
\left\langle s \mu_f ; \vec{p}_2|\mathcal{T}| \vec{p}_i ; s_i \mu_i\right\rangle= & \sum_{l_2, m_f} \sum_{l_i, m_i} \sum_{j, j^{\prime}} C\left(l_2 s j^{\prime} ; m_f \mu_f m^{\prime}\right) C\left(l_i s_i j ; m_i \mu_i m\right) \\
&\times Y_{l_{2}m_f}\left(\hat{{\mathbf{p}}}_2\right)\left\langle\left(l_2 s\right) j^{\prime} m^{\prime}|\mathcal{T}|\left(l_i s_i\right) j m\right\rangle Y_{l_i m_i}^*\left(\hat{{\mathbf{p}}}_i\right),
\end{aligned}
\end{equation}
where $C(j_1,j_2j;m_1m_2m)$ denote the Clebsch-Gordan coefficients in the notation of Rose \cite{rose} and the rotational invariance implies
$$
\left\langle\left(l_2 s\right) j^{\prime} m^{\prime}|\mathcal{T}|\left(l_i s_i\right) j m\right\rangle=\delta_{j j'} \delta_{m m^{\prime}} \mathcal{T}_{l_2 s; l_i s_i}^j
$$
for the energy-dependent partial-wave amplitudes, $\mathcal{T}_{l_2 s; l_i s_i}^j$. If $A^k_q$ is the $q$th component of an irreducible (spherical) tensor $A^k_q$ of rank $k$ \cite{varsh} with $q=-k,\ldots,+k$ and $B^{k'}_{q'}$ the $q'$th component of an irreducible tensor of rank $k'$ with $q'=-k',\ldots, +k'$, we use the short hand notation
$$
\left(A^k \otimes B^{k^{\prime}}\right)_Q^K=\sum_{q=-k}^k C\left(k k^{\prime} K, q q^{\prime} Q\right) A_q^k\, B_{q'}^{k^{\prime}}
$$
to denote the $Q$th component of an irreducible tensor of rank $K$ constructed out of the two irreducible tensors $A_q^k$ and $B_{q'}^{k'}$. Using Racah techniques, eq. (\ref{partial}) may be rewritten as \cite{fesh}
\begin{equation}
\begin{aligned}
\label{T}
\left\langle s \mu_f ; \vec{p}_2|\mathcal{T}| \vec{p}_i ; s_i \mu_i\right\rangle= & \sum_{l_2, l_i, j, \lambda}(-1)^{l_i+s_i+l_2-j} W\left(s_i l_i s l_2 ; j \lambda\right)[j]^2[\lambda]\left[s\right]^{-1} \mathcal{T}_{l_2 s;l_i s_i}^j \\
&\times C\left(s_i \lambda s ; \mu_i \mu \mu_f\right)(-1)^\mu\left(Y_{l_2}\left(\hat{{\mathbf{p}}}_2\right) \otimes Y_{l_i}\left(\hat{{\mathbf{p}}}_i\right)\right)_{-\mu}^\lambda,
\end{aligned}
\end{equation}
where the $W()$s are the Racah coefficients \cite{varsh}. Also, in eq. (\ref{T}) we use the shorthand notation
$$
[j]\equiv \sqrt{(2j+1)}
$$
and as an exception (to be consistent with the notation used everywhere else in the literature), only for the spherical harmonics, we show $l_2$ and $l_i$ as suffixes even though they are the ``ranks'' of the (irreducible) spherical harmonics.
We now express $\mathcal{M}$ as
\begin{equation}
\label{coll}
\mathcal{M}=\sum_{s, \mu_f} \sum_{s_i, \mu_i} \sqrt{\frac{2 \pi D}{v}}\left|s\mu_f\right\rangle\left\langle s \mu_f ; \overrightarrow{p_2}|\mathcal{T}| \overrightarrow{p_i} ; s_i \mu_i\right\rangle\left\langle s_i \mu_i\right|
\end{equation}
and use eq. (\ref{T}) for the matrix elements. Since $\left|s \mu_f\right\rangle$ transforms under rotations as an irreducible tensor, $K_{\mu_f}^{s}$, of rank $s$; while $\left\langle s_i \mu_i\right|$ does so like $(i)^{2 \mu_i} B_{-\mu_i}^{s_i}$, an irreducible tensor of rank $s_i$ \cite{sch}, so that
\begin{equation}
\label{proj}
\left|s \mu_f\right\rangle\left\langle s_i \mu_i\right|=\sum_{\lambda=|s_i-s|}^{(s_i+s)} C\left(s s_i \lambda ; \mu_f-\mu_i \mu\right)(-1)^{\mu_i}\left(K^{s} \otimes B^{s_i}\right)_\mu^\lambda.
\end{equation}
This allows us to define irreducible spin transition tensors $\mathcal{S}_\mu^\lambda$ connecting the spin spaces of $s_i$ and $s$ through
\begin{equation}
\label{proj-channel}
\mathcal{S}_\mu^\lambda\left(s, s_i\right)=(-1)^{s_i}\left[s\right]\left(K^{s} \otimes B^{s_i}\right)_\mu^\lambda,
\end{equation}
where $\lambda$ ranges from $\left|s-s_i\right|$ to $\left(s+s_i\right)$. The collision matrix, eq. (\ref{coll}), can be written on using eqs. (\ref{proj}) and (\ref{proj-channel}), in the simple {\em invariant} form \cite{npa,gr-msv}
\begin{equation}
\label{invariant}
\mathcal{M}=\sum_{s, s_i, \lambda}\left(\mathcal{S}^\lambda\left(s, s_i\right) \cdot \mathcal{T}^\lambda\left(s, s_i\right)\right)\equiv\sum_{s, s_i, \lambda}\ \sum_{\mu=-\lambda}^\lambda(-1)^\mu\,\mathcal{S}^\lambda_\mu\left(s, s_i\right)\,\mathcal{T}^\lambda_{-\mu}\left(s, s_i\right) ,
\end{equation}
where
$$
\begin{aligned}
\mathcal{T}_{-\mu}^\lambda\left(s, s_i\right)= \sum_{l_2, l_i, j}(-1)^{l_i+s_i+l_2-j} W\left(s_i l_i s l_2 ; j \lambda\right) \mathcal{M}_{l_2 s ; l_i s_i}^j[j]^2\left[s\right]^{-1} 
\left(Y_{l_2}\left(\hat{{\mathbf{p}}}_2\right) \otimes Y_{l_i}\left(\hat{{\mathbf{p}}}_i\right)\right)_{-\mu}^\lambda
\end{aligned}
$$
are the irreducible tensor reaction amplitudes in the channel-spin space and
$$
\mathcal{M}_{l_2 s ; l_i s_i}^j=\sqrt{\frac{2 \pi D}{v}} \mathcal{T}_{l_2 s; l_i s_i}^j.
$$
Specializing the above formalism for the case of $\overline{p} p\rightarrow \bar{\Xi}^{+} \Xi^{*-}$ involves choosing the spins to be $s_a=s_A=1/2$ with $s_i=0,1$ and if $s_b=s_B=3/2$,  $s=0,1,2,3$, while if $s_b=s_B=1/2$,  $s=0,1$. Note that invariance under parity implies 
$$
(-1)^{l_i}=(-1)^{l_2}.
$$
Also, if the $\Xi$s are produced in a s-wave, $l_2=0$.

\subsection{$\overline{p} p \rightarrow \bar{\Xi}^{+} \Lambda K^{-}$}
Similar to the formalism detailed in the previous section and following \cite{gr-pnd}, the reaction matrix for $\overline{p} p \rightarrow \bar{\Xi}^{+} \Lambda K^{-}$, can be written as
$$
\mathcal{M}=\sum_{s_f}\sum_{s_i=0}^1 \sum_{\lambda=\left|s_f-s_i\right|}^{s_f+s_i}\left(S^\lambda\left(s_f, s_i\right) \cdot \mathcal{M}^\lambda\left(s_f, s_i\right)\right),
$$
where $s_f=0,1$ if $\Xi^*$ has spin-1/2, while $s_f=1,2$ if $\Xi^*$ has spin-3/2 and the irreducible amplitudes 
\begin{align}
\mathcal{M}_\mu^\lambda\left(s_f, s_i\right)= \sum_{l_f, l, L_f, j_f, j, l_i} g_\alpha \mathcal{M}_{l\left(l_f s_f\right) j_f ; l_i s_i}^j 
\left(\left(Y_{l_f}\left(\hat{\mathbf{p}}_f\right) \otimes Y_l(\hat{\mathbf{q}})\right)^{L_f} \otimes Y_{l_i}\left(\hat{\mathbf{p}}_i\right)\right)_\mu^\lambda
\end{align}
are written in terms of the partial-wave amplitudes, $\mathcal{M}_{l\left(l_f s_f\right) j_f ; l_i s_i}^j$, with $g_\alpha$ being geometrical factors given by
$$
\begin{aligned}
g_\alpha= & (4 \pi)^3(i)^{l_i-l-l_f}(-1)^{l+l_f+l_i-j+s_i}[j]^2\left[j_f\right]\left[L_f\right] \\
& \times\left[s_f\right]^{-1} W\left(l_i s_i L_f s_f ; j \lambda\right) W\left(s_f l_f j l ; j_f L_f\right),
\end{aligned}
$$
where $\alpha$ denotes, collectively, $\alpha=\left\{l, l_f, L_f, s_f, j_f, j,l_i,s_i,\lambda\right\}$.
In the above equation, $\mathbf{p}_i=p_i \hat{\mathbf{p}}_i, \mathbf{p}_f=p_f \hat{\mathbf{p}}_f$ denote, respectively, the initial and final momenta associated with the relative motion of the $\Xi^*$ and $\Lambda$, while $\mathbf{q}=q \hat{\mathbf{q}}$ denotes the $K^-$ momentum in c.m. At threshold, we assume $l,l_f=0,1$. 
If we denote $l'$ to be the relative angular momentum between the $\Lambda$ and $K^-$, then $l'=0,1$ if $\Xi^*$ has spin-1/2, while $l'=1,2$ if $\Xi^*$ has spin-3/2.

\section{Relating the partial-waves of $\bar{p} p \rightarrow \bar{\Xi}^{+} \Xi^{*-}$ to that of $\overline{p} p \rightarrow \bar{\Xi}^{+} \Lambda K^{-}$}
If $\mathbf{p}_1$ and $\mathbf{p}_2$   denote the c.m. momenta of the 
$\Lambda$ and $\Xi^+$ respectively,
then 
$$
\mathbf{p}_f=\frac{M_\Xi\mathbf{p}_1-M_\Lambda\mathbf{p}_2}{M_\Lambda+M_\Xi},
$$
where $M_\Xi$ and $M_\Lambda$ are the masses of $\Xi^+$ and $\Lambda$ respectively. If $\mathbf{q}'$ is the relative momentum between the $K^-$ and the $\Lambda$, with $\mathbf{q}$ being the c.m. momentum of the $K^-$, then
$$
\mathbf{q}'=\alpha \mathbf{p}_f+\beta \mathbf{q},
$$
where 
$$
\alpha=\frac{-M_K}{M_\Lambda+M_K},
$$
where $M_K$ is the mass of $K^-$ and
$$
\beta=\frac{M_\Lambda^2+M_\Lambda M_K+M_\Lambda M_\Xi}{(M_\Lambda+M_\Xi)(M_\Lambda+M_K)}
$$
and
$$
\mathbf{p}_2=-\mathbf{p}_f-\frac{M_\Xi}{M_\Lambda+M_\Xi}{\mathbf{q}}.
$$
We now use eq. (51) in \cite{venkataraya} to relate the angular momenta $l',l_2$ for $\overline{p}p \rightarrow \bar{\Xi}^{+} \Xi^{*-}$ with the angular momenta $l_f,l$ for $\overline{p}p \rightarrow \bar{\Xi}^{+} \Lambda K^{-}$, viz.,
\begin{equation}
\begin{aligned}
\label{venkat}
\left(Y_{l^{\prime}}\left(\hat{\mathbf{q}}^{\prime}\right) \otimes Y_{l_2}\left(\hat{\mathbf{p}}_2\right)\right)_{M_f}^{L_f}&=\sum_{L=0}^{l'}\sum_{L'=0}^{l_2}\,\sum_{l,l_f=0}^1 F\, C\left(L L^{\prime} l ; 000\right)
 C\left(l^{\prime}-L, l_2-L^{\prime}, l_f ; 000\right)\\ &\times
\left\{\begin{array}{ccc}
L & L^{\prime} & l \\
l^{\prime}-L & l_2-L^{\prime} & l_f \\
l^{\prime} & l_2 & L_f
\end{array}\right\}\left(Y_l(\hat{\mathbf{q}}) \otimes Y_{l_f}\left(\hat{\mathbf{p}}_f\right)\right)_{M_f}^{L_f},
\end{aligned}
\end{equation}
where $F$ is a factor whose expression can be found in \cite{venkataraya} and irrelevant for our subsequent conclusions.
Note that eq. (49) in \cite{venkataraya} would be modified to
$$
\begin{gathered}
Y_{l_2 m_2}\left(\hat{\mathbf{p}}_2\right)=\sqrt{4 \pi}(-1)^{l_2} \sum_{L^{\prime}=0}^{l_2} \frac{1}{\left[L^{\prime}\right]}\binom{2 l_2+1}{2 L^{\prime}}^{\frac{1}{2}} 
\frac{(q\delta )^{L^{\prime}}\left(p_f\right)^{l_2-L^{\prime}}}{(2)^{L^{\prime}} p_2^{l_2}}\left(Y_{L^{\prime}}(\hat{\mathbf{q}}) \otimes Y_{l_2-L^{\prime}}\left(\hat{\mathbf{p}}_f\right)\right)_{m_2}^{l_2},
\end{gathered}
$$
where $\delta=(2M_\Xi)/(M_\Lambda+M_\Xi)$.
\section{Results and Conclusion}
From eq.(\ref{venkat}) and using the properties of the Clebsch-Gordan coefficients and the 9-j symbol, we clearly see that the partial-wave amplitude corresponding to $l_2=0$ {\em cannot} contribute to $l=l_f=1$, if $\Xi^*$ has spin-1/2 (in which case, $l'=0,1$) but it {\em can} contribute {\em only} to a $\Xi^*$ which has a spin-3/2, because $l'=2$ is an allowed angular momentum quantum number in this case and hence the partial-wave amplitude in eq. (7) can have $l=l_f=1$. 

Any observable measured for the reaction $\overline{p} p \rightarrow \bar{\Xi}^{+} \Lambda K^{-}$ will be a bilinear in the reaction amplitudes given by eq. (7). That is, they will involve real and imaginary parts of $\mathcal{M}^\lambda \mathcal{M}^{\lambda'*}$. If any of these observables, say, $O(\hat{\mathbf{q}},\hat{\mathbf{p}}_f)$, is measured as a function of $\hat{\mathbf{q}}$ and $\hat{\mathbf{p}}_f$, then
\begin{align}
\int O(\hat{\mathbf{q}},\hat{\mathbf{p}}_f)\, Y_{2m}(\hat{\mathbf{q}}) Y_{2m'}(\hat{\mathbf{p}}_f)\, d\Omega_{\hat{\mathbf{p}}_f}d\Omega_{\hat{\mathbf{q}}}
\end{align}
for any $m,m'$ would be zero if $\Xi^*$ had spin-1/2. This is because, the only way in which the observable $O(\hat{\mathbf{q}},\hat{\mathbf{p}}_f)$ could contain a $Y_{2m}(\hat{\mathbf{q}})$ and a $ Y_{2m'}(\hat{\mathbf{p}}_f)$ so that the orthogonality of the $Y_{lm}$s will not make the integral in eq. (2) zero is for it to contain an $l=1$ and an $l_f=1$ inside it. This is possible only if $\Xi^*$ has spin-3/2, as we saw earlier.
\begin{acknowledgments}
    The author acknowledges with gratitude the financial support received under the DST-SERB grant, CRG/2021/000435.
\end{acknowledgments}

\end{document}